\documentclass[aps,prc,twocolumn,floatfix,inputenc]{revtex4}
\usepackage{graphicx,color}
\usepackage{amsmath}
\usepackage{amssymb}
\usepackage{bm}
\usepackage{amsfonts}

\usepackage{graphicx}

\begin{document}
\title{Final state interaction in the pn and nn decay channels of  $^4_\Lambda$He}

\author{C. A. Bertulani} \email{carlos.bertulani@tamuc.edu} 
\affiliation{Department of Physics and Astronomy, Texas A\&M University-Commerce, Commerce, Texas 75429, USA}
\author{R. Lobato}\email{ronaldo.lobato@tamuc.edu}
\affiliation{Department of Physics and Astronomy, Texas A\&M University-Commerce, Commerce, Texas 75429, USA}
\begin{abstract}
We study the effects of final state interactions in the non-mesonic weak decay $\Lambda N \rightarrow nN$  (n is a neutron and N is either a neutron or a proton) of the hypernucleus $_\Lambda^4$He. Using a three-body model the effects of distortion of the interaction of the emitted nucleon pair with the residual nucleus is considered. We also study the influence of the final state interaction between the emitted nucleons using the Migdal-Watson model. The effect of spin symmetries in the final state of the pair is also considered. Based on our calculations, we conclude that final state interactions play a minor role in the kinetic energy spectrum of the emitted nucleon pair.
\end{abstract}

\maketitle

\hfil {\it To the memory of M.S. Hussein}

\section{Introduction}
\label{intro}
Electrons and nucleons  make up most of the visible matter in the universe, with the nucleons being composed of  up and down valence quarks. In a hyperon, a strange quark replaces one of up or down quark. A hypernucleus is created by a reaction such as K + A $\rightarrow$ $\pi$ +  hypernucleus, with the outcome that a hyperon is implanted as an impurity within the nucleus A. The $\Lambda$-particle, with an up-down-strange particle composition, has isospin and spin 0, and its implantation forms a common kind of hypernucleus, first observed in 1952 \cite{DP53}.  In free space, the lambda decays with a total decay rate $\Gamma = 2.5 \times 10^{-6}$ eV, and a corresponding mean lifetime of $263\pm 2$ ps \cite{Ams08}, into p + $\pi^-$ (64.1\%) or n + $\pi^0$ (35.7\%).   

Due to Pauli blocking, the mesonic channel $\Lambda \rightarrow \pi + N$ is strongly inhibited in a heavy (e.g., $^{12}$C or heavier) nucleus and the non-mesonic weak decay (NMWD) $\Lambda N \rightarrow nN$ becomes a dominant channel. We expect it to be negligible for the case of the $^4$He hypernucleus studied in this work. The process can be induced by a neutron ($N=n$) or a proton ($N=p$) with respective widths $\Gamma_n$ and $\Gamma_p$. The non-mesonic decay process $\Lambda N \rightarrow NN$ can occur due to the combination of a weak vertex and a strong vertex in the process and are dominated by the proton-induced reaction $\Lambda p \rightarrow np$ and neutron-induced reaction. In the nuclear medium these widths are modified  because the nucleon and the hyperon can move  in their respective mean fields arising from the NN and N$\Lambda$ interactions. The fact that three-body non-mesonic decay modes are negligible, but two-body non-mesonic decay comprise nearly 20\% of the events, implies that the three-body decay modes probably do not play an important role in the emission spectra and would fail to explain the observed discrepancies between a two-body decay calculation and the data. In this work, we plan to ascertain if this assertion is valid.   

One of the interests in the study of non-mesonic decays of light hypernuclei such as $_\Lambda^4$He is because they are expected to be the only feasible method to study the $\Delta S = 1$ weak baryon-baryon interaction with the weak interaction Hamiltonian due to the W-exchange given by \cite{Oku82}
\begin{equation}
H_{weak}={G_F \over \sqrt{2}} \sin \theta_c \cos \theta_c \left[ \bar{u} \gamma_\mu (1-\gamma_5)s\bar{d}\gamma^\mu(1-\gamma_5)u\right],
\end{equation}
where $u$, $s$, $d$ are quark spinors. While this Hamiltonian does not favor $\Delta T = 1/2$ over $\Delta T =3/2$ weak transitions, the  $\Delta T = 1/2$ weak transitions are found to dominate in experiments.  This is not a well understood fundamental phenomenon. 

For $\Delta T = 1/2$ the effective weak Hamiltonian can be written in terms of a nucleon, $\Psi_N$, a lambda, $\Psi_\Lambda$, and a  pion field, $\mbox{\boldmath$\phi$}_\pi$. That is,
\begin{equation}
H_{weak}=-G_F m_\pi^2 \bar{\Psi}_N (A_\pi + B_\pi \gamma_5) \mbox{\boldmath$\phi$}_\pi \cdot \mbox{\boldmath$\tau$}\Psi_\Lambda\left(\begin{array}{c}0 \\1\end{array}\right),
\end{equation}
where  the weak coupling constant is $G_Fm^2_\pi = 2.21 \times 10^{-7}$,  and the constants  $A_\pi = 1.05$ and $B_\pi = - 7.15$ are adjusted to reproduce the free lambda decay. The isospin spurion $\left(\begin{array}{c}0 \\1\end{array}\right)$ enforces $\Delta T = 1/2$ \cite{Oku82,Isg90}.

Hypernuclei also possess properties of relevance for the study of nuclear structure, medium modification of particle properties, and stellar processes. For example, the ground state of helium-4 with an implanted lambda, here denoted as $^4_\Lambda$He, must be similar to $^4$He  and must be stable, except that an eventual weak decay of the lambda occurs. Therefore, precise measurements of  the mass and the excitation energies of a hypernucleus reveals the medium properties of the hyperon-nucleon  interaction  of relevance for the understanding of high density nuclear matter occurring in stellar environments such as within neutron stars \cite{GHM16}. 

The induced decay of  a hypernucleus into two emitted nucleons, usually termed non-mesonic weak decay (NMWD),  ($\Lambda$N $\rightarrow$ nN),
is evidently a complex problem, involving more than one particle. To qualitatively describe the decay, one can resort to very simple
models for the decay width, based on, e.g., the mean field description of the bound system and free nucleons (e.g., plane waves)
in the final channel. But to account for the effect of final state interactions (FSI), such as the interaction of the emerging
nucleons with the field of the residual nucleus and the pn and nn interactions, one has to go well beyond the above
simple model. 

The aforementioned plane wave model fails rather drastically in describing the
nn channel, while it works adequately for the pn channel, as shown, e.g., in Refs. \cite{Bar08,Bau09}. This was traced to the fact that the p in the pn channel
suffers a push owing to the Coulomb field of the residual nucleus, which
renders the pn correlation effects rather weak.
The plane wave description, which exhibits a peaking of the pn spectrum as a function of the sum kinetic energies
of the two nucleons comes out at the right place with the data. Thus two final state interactions seem to cancel
each other in the pn channel. This feature is not reflected in the nn channel, since the  n-residual nucleus final state interaction
is the same for both neutrons and thus the nn correlation survives. The comparison of the simple plane wave model
with the data clearly shows the need for this correlation: the peaking in the calculation is way to high. The data show a
peaking at much lower sum kinetic energies. Needless to say that besides the decay modes
$\Lambda$n $\rightarrow$ nn and  $\Lambda$p $\rightarrow$ np  there are also  two-nucleon induced modes
$\Lambda$NN $\rightarrow$ nNN, which is an evident three-body problem. 

This work is build up on the ideas explained in some of M.S. Hussein's publications with his collaborators \cite{Bar08,Bau09,Bau09b,KGH10,Baue10}. We extend their work an we concentrate our efforts in exploiting aspects of the three-body problem in the case of $\Lambda$NN $\rightarrow$ nNN and the final state interaction involving the residual nucleus and the nucleons. The basic goals of this work is to show  that final state interactions can be treated in different ways, leading to different predictions for the kinetic energy sum of the pn and nn spectra. Although we compare our results with the specific case of the NMWD of the $^4_\Lambda$He nucleus, our ideas are very general and can be applied to decay of any other hypernucleus.

\section{The three-body final state interaction}
In this section we consider a pn emitted pair for the sake of simplifying the notation, although all developments can also be used for nn pairs.  In order to take into account the final state interaction we use a three-body description based on the Faddeev
equation for the full final state wave function for the pn+(A-2) system, $\Psi_f^{(-)} ({\bf p}_n, {\bf p}_p, {\bf P}_{A-2})$,
\begin{eqnarray}
\left[ E-H_0-U^\dagger_{pA-2}-U^\dagger_{nA-2}-V_{pn}\right]&\Psi_f^{(-)} ({\bf p}_n, {\bf p}_p, {\bf P}_{A-2})
\nonumber\\ &=0,\label{Htot}%
\end{eqnarray}
where $H_0$ is a sum of the kinetic energy operator of the two emerging nucleons, $K_p$, $K_n$,  and of the residual nucleus, plus the intrinsic Hamiltonian of the residual nucleus, $h_{A-2}$. The quantity
\begin{equation}
h_{A-2}+U_{pA-2}+U_{nA-2}+V_{pn},\label{Htot2}%
\end{equation}
describes the bound three-body system immediately after the decay of the $\Lambda$ hyperon (the two nucleons
are still inside the  nucleus) and, by definition, $V_{pn}$ is real.

The direct solution of the three-body Schr\"odinger equation, Eq. (\ref{Htot}), is not possible when only two-body
interactions are present, as is our case.  On the other hand, the Faddeev decomposition allows a legitimate
method of finding the solution. For this purpose, we first write the formal solution of Eq. (\ref{Htot}), namely the
Lippmann-Schwinger equation for the full final state wave function, as
\begin{eqnarray}
\Psi_f^{(-)} &=& \Phi_f^{(-)}+( E^{(-)}-H_0)^{-1}\nonumber\\
&\times& \left(U_{pA-2}^\dagger+U_{nA-2}^\dagger+V_{pn}\right)\Psi_f^{(-)},\label{Psitot}
\end{eqnarray}
where the inhomogeneous solution  $\Phi_f^{(-)}$ corresponds to the two outgoing nucleons and the left over residual
A-2 nucleus. It is a solution of the equation without the n(A-2) and p(A-2) potentials.
Following the usual procedure to solve Faddeev-like equations, we write
\begin{equation}
\Psi^{(-)} = \Phi_{fpn}^{(-)}+\Psi_{pA}^{(-)}+\Psi_{nA}^{(-)}+\Psi_{pn}^{(-)},\label{Psifad}
\end{equation}
where the Faddeev components are given by the solution of coupled Schr\"odinger
equations. Consider the one accounting for the relative motion of the pair, $\Psi_{pn}$,
\begin{equation}
\Psi^{(-)}_{pn} = ( E^{(-)}-H_0)^{-1}V_{pn}\left[
\Phi_{fpn}^{(-)}+\Psi_{pA}^{(-)}+\Psi_{nA}^{(-)}+\Psi_{pn}^{(-)}\right].\label{Psifad2}
\end{equation}
In Eqs. (\ref{Htot2},\ref{Psifad}-\ref{Psifad2}), the ``free" three-body Green's function, $( E^{(-)} - H_0 )^{-1}$, is an outgoing one with $E^{(-)}  = E -i\epsilon$.

The next step is to formally solve for $\Psi^{(-)}_{pn}$ by writing
\begin{eqnarray}
\left[1-( E^{(-)}-H_0)^{-1}V_{pn}\right] \Psi^{(-)}_{pn} &=& ( E^{(-)}-H_0)^{-1}V_{pn} \\
&\times& \left[\Phi_{fpn}^{(-)}+\Psi_{pA}^{(-)}+\Psi_{nA}^{(-)}\right],\nonumber\label{Psifad3}
\end{eqnarray}
or, with the identity $[ 1 - ( E^{(-)} - H_0 )^{-1} V_{pn} ]^{-1} = ( E^{(-)} - H_0 )^{-1} t^\dagger_{pn}$,
\begin{eqnarray}
\Psi^{(-)}_{pn} &=& ( E^{(-)}-H_0)^{-1}t^\dagger_{pn}
\Phi_{fpn}^{(-)}\nonumber \\
&+&( E^{(-)}-H_0)^{-1}t^\dagger_{pn}\left( \Psi_{pA}^{(-)}+\Psi_{nA}^{(-)}\right),\label{Psifad4}
\end{eqnarray}
where $t_{pn}$ is the two-nucleon $t$-matrix operating in the three-body Hilbert space.

Eqs. (\ref{Psifad}-\ref{Psifad4}) are the Faddeev equations to be solved to determine the there-body final state wavefunction. Once
they are solved, the full wave function is constructed by their
sum. In most applications one component dominates. In the simple
model mentioned above, the full wave function is replaced by the
single particle wave function with no nucleon-nucleon
correlations. The Faddeev equations above contain the single-particle aspect through the potentials $U_{pA-2}$  and $U_{nA-2}$
and the full NN correlations in the final state. In this section, we do not
consider correlations in the bound state wave function besides the
pn one. Effects of symmetry  and their consequence for correlation functions a pair of identical nucleons have been published in the literature, e.g., in Ref. \cite{BHV08}.

To pin down the driving physics on the $\Psi^{(-)}_{ pA} + \Psi^{(-)}_{nA}$ part of the three-body wave
function, we resort to the limit that $V_{pn}$ vanishes and the
recoiling nucleus has an infinite mass. In this situation the full
three-body wave function is just $\chi^{(-)}({\bf p}_p)
\chi^{(-)}({\bf p}_n)$, and the sum of Faddeev components
$\Psi^{(-)}_{ pA} + \Psi^{(-)}_{nA}$ is the product minus the
inhomogeneous form (\ref{Psitot}). This leading order term would
gain multiple scattering corrections by increasing the importance
of $V_{pn}$ and two-nucleon correlations would start to contribute to
these two Faddeev components. 

Based on the above discussion, we now approximate the less important summed wave
components  $\Psi^{(-)}_{ pA} + \Psi^{(-)}_{nA}$  by a product of
distorted waves for p and n, minus the inhomogeneous solution, namely,
$\chi^{(-)}({\bf p}_p) \chi^{(-)}({\bf p}_n) - \Phi_f^{(-)}$. 
This gives,
\begin{eqnarray}
\Psi^{(-)}_{pn} &=& ( E^{(-)}-H_0)^{-1}t^\dagger_{pn}
\Phi_{fpn}^{(-)} \\
&+&( E^{(-)}-H_0)^{-1}t^\dagger_{pn}\left[ \chi^{(-)}({\bf p}_p) \chi^{(-)}({\bf p}_n) - \Phi_f^{(-)}\right],\nonumber \label{Psifad3}%
\end{eqnarray}
and accordingly,
\begin{equation}
\Psi^{(-)}_{pn} = ( E^{(-)}-H_0)^{-1}t^\dagger_{pn}
\chi^{(-)}({\bf p}_p) \chi^{(-)}({\bf p}_n). \label{Psifad5}%
\end{equation}
The full final state three-body wavefunction of Eq. (\ref{Htot2}) can now be expressed as
\begin{eqnarray}
\Psi_f^{(-)} &=& \Phi_f^{(-)}+\Psi_{pA}^{(-)} + \Psi_{nA}^{(-)}\nonumber\\
&+&( E^{(-)}-H_0)^{-1}t^\dagger_{pn}
\chi^{(-)}({\bf p}_p) \chi^{(-)}({\bf p}_n).\label{Psitot2}%
\end{eqnarray}
But,
\begin{equation}
\Psi^{(-)}_{ pA} + \Psi^{(-)}_{nA}=\chi^{(-)}({\bf p}_p) \chi^{(-)}({\bf p}_n) - \Phi_f^{(-)}, \label{Psifad5}%
\end{equation}
and we finally obtain the desired approximation for the final state wavefunction
\begin{equation}
\Psi^{(-)}_{f} = \left[ 1 +( E^{(-)}-H_0)^{-1}t^\dagger_{pn}\right]
\chi^{(-)}({\bf p}_p) \chi^{(-)}({\bf p}_n)\Psi_{A-2}, \label{Psifad5}%
\end{equation}
which will be used as the starting point of our calculations. 

It is important to describe how the hypernucleus decay width depends on the approximations used for the wavefucntions and how they relate to the final momenta and energies of the emitted pair. We demonstrate this connection in the coming section.

\section{Decay width}
The weak non-mesonic decay width is given by the Fermi Golden Rule. The non-mesonic weak-decay rate of a hypernucleus from its ground state $i$  to the residual nucleus ground state  and two free nucleons $pn$\footnote{And similarly for nn pairs.} with total spin $S$, total kinetic energy $E_{pn, cm} = E_n + E_p$ and relative energy $E_{pn, rel}$, is given by
\begin{eqnarray}
\Gamma&=&\frac{2\pi}{\hbar}\sum\int \left|\langle\Psi_f^{(-)} ({\bf p}_n, {\bf p}_p, {\bf P}_{A-2})\left|V_{weak}\right|\Psi_{i,A}(J_1M_1)
\rangle\right|^2\nonumber\\
&\times&
\delta\left(E_{pn,cm}+E_{pn,rel}-\Delta_n\right)\frac{d{\bf
p}_p}{(2\pi)^3}\frac{d{\bf p}_n}{(2\pi)^3} ,
\end{eqnarray}
where $\Delta_n$ includes the energies for different final states with the same spin $J_f$. Here we consider ground-state transitions so that
\begin{eqnarray}
\Delta_n = M_\Lambda - M =176 \ {\rm MeV}.
\end{eqnarray}

In the calculation of the width we need the complex adjoint of  $\Psi_f^{(-)}$, namely,
\begin{eqnarray}
\Psi^{(-)*}_{f} &=& \chi^{(-)*}({\bf p}_p) \chi^{(-)*}({\bf p}_n)\nonumber \\
&\times&\Psi_{A-2}\left[ 1 +t_{pn}( E^{(+)}-H_0)^{-1}\right]. \label{Psifad6}
\end{eqnarray}
Thus, in a prior-form of the final state interaction for the decay
amplitude, i.e., when the distortion of the two-nucleon final
state by the NN potential occurs just after the weak process, the
individual nucleon optical distortion prevails, and the width
becomes
\begin{eqnarray}
\Gamma&=&\frac{2\pi}{\hbar}\sum\int \Bigg|\Big\langle\chi^{(-)}({\bf
p}_p)\chi^{(-)}({\bf
p}_n),\Psi_{A-2}(SM_SJ_FM_F)\Big| \nonumber \\
&\times&\left[1+t_{pn}\left(\Delta_n-\frac{\hbar^2{\bf
P}^2_{pn,cm}}{4m_N}\right)\left(\Delta_n-H_0+i\varepsilon\right)^{-1}\right]
\nonumber \\  &\times& V_{weak}\Big|\Psi_{A}(J_IM_I)\Big\rangle\Bigg|^2
\delta\left(E_{pn,cm}+E_{pn,rel}-\Delta_n\right)\nonumber \\
&\times&\frac{d{\bf
p}_p}{(2\pi)^3}\frac{d{\bf p}_n}{(2\pi)^3} \ .
\end{eqnarray}

To derive the above formula one has to remind that the energy
of the three-body system in the center of mass is $E=\Delta_n$, as
constrained by energy conservation. The three-body resolvent
carries this energy minus the kinetic energy operator. The
neutron-proton T-matrix is calculated within the three-body
background where the remaining nucleus is a spectator, then the
two-body energy has its argument  given by
$\Delta_n-{\hbar^2{\bf P}^2_{pn,cm}}/{4m_N}$, where the pn
center of mass kinetic energy operator is subtracted from the
total energy of the three-body system. It is disregarded the
dependence on spin of the distorted waves of the nucleons, and
therefore the total $|J_FM_F\rangle$ state is the sum of the total
np spin and the spin of the remaining nucleus.

Rearranging the terms, 
\begin{eqnarray}
\Gamma&=&\frac{2\pi}{\hbar}\sum\int \Bigg|\Big\langle\chi^{(-)}({\bf
p}_p)\chi^{(-)}({\bf p}_n)\Big|\nonumber \\
&\times&\left[1+t_{pn}^{S}\left(\Delta_n-\frac{\hbar^2{\bf
P}^2_{pn,cm}}{4m_N}\right)\left(\Delta_n-H_0+i\varepsilon\right)^{-1}\right]
\nonumber \\
&\times&
 %\langle
\Psi_{A-2}(SM_SJ_FM_F)\Big|V_{weak}\Big|\Psi_{A}(J_IM_I)\rangle\Bigg|^2
\nonumber\\
&\times&\delta\left(E_{pn,cm}+E_{pn,rel}-\Delta_n\right)\frac{d{\bf
p}_p}{(2\pi)^3}\frac{d{\bf p}_n}{(2\pi)^3},
\end{eqnarray}
and inserting unit resolutions in the above equation one has that
\begin{eqnarray}
\Gamma&=&\frac{2\pi}{\hbar}\sum\int \frac{d{\bf
p}_p}{(2\pi)^3}\frac{d{\bf p}_n}{(2\pi)^3} \Big|\int d{\bf p}^\prime_p
d{\bf p}^\prime_n d{\bf p}^{\prime\prime}_p d{\bf
p}^{\prime\prime}_n\nonumber \\
&\times&\langle\chi^{(-)}({\bf p}_p)|{\bf
p}^\prime_p\rangle\langle\chi^{(-)}({\bf p}_n)|{\bf
p}^\prime_n\rangle K^{(S)}({\bf p}^\prime_p,{\bf p}^\prime_n;{\bf
p}^{\prime\prime}_p,{\bf p}^{\prime\prime}_n)
\nonumber \\
&\times& \langle {\bf p}^{\prime\prime}_n,{\bf
p}^{\prime\prime}_n,
\Psi_{A-2}(SM_SJ_FM_F)|V_{weak}|\Psi_{A}(J_IM_I)\rangle|^2\nonumber \\
&\times&\delta\left(E_{pn,cm}+E_{pn,rel}-\Delta_n\right)  \label{width} ,
\end{eqnarray}
where the kernel containing the two-nucleon scattering matrix is
evaluated for a contact s-wave interaction, giving
\begin{eqnarray}
&&K^{(S)}({\bf p}^\prime_p,{\bf p}^\prime_n;{\bf
p}^{\prime\prime}_p,{\bf p}^{\prime\prime}_n)=\delta({\bf
p}^\prime_p-{\bf p}^{\prime\prime}_p)\delta({\bf p}^\prime_n-{\bf
p}^{\prime\prime}_n)\nonumber \\ &+&\delta\left({\bf
P}^{\prime\prime}_{pn,cm}-{\bf
P}^\prime_{pn,cm}\right)\Bigg\{ 1 + \tau_{pn}^{S}\left(\Delta_n-\frac{{\hbar^2{\bf
P}^\prime}^2_{pn,cm}}{4m_N}\right)  \nonumber \\ &\times&\left[\Delta_n-\frac{({\hbar{\bf
P}^{\prime\prime}_{pn,cm}})^2 }{4m_N}-\frac{({\hbar{\bf
P}^{\prime\prime}_{pn,rel}})^2}{m_N} +i\varepsilon\right]^{-1} \Bigg\} ,
\end{eqnarray}
where the Jacobi momenta are
\begin{eqnarray}
{\bf P}^{\prime}_{pn,cm}={\bf p}^\prime_p+{\bf p}^{\prime}_n
~~~,~~ {\bf P}^{\prime}_{pn,rel}=\frac12({\bf p}^\prime_p-{\bf
p}^{\prime}_n) \ ,
\end{eqnarray}
and analogous definitions are used for double primed quantities. 

The nucleon-nucleon amplitude is given by
\begin{eqnarray}
&&\tau_{pn}^{S}\left(E\right)=\nonumber \\&&\left[\int d{\bf
p}\left[\left(k^2\cot^2\delta-{\bf p}^2\right)^{-1}-\left(E-{\bf
p}^2+i\varepsilon\right)^{-1}\right]\right]^{-1} \nonumber \\ &=&
\left[\alpha+i4\pi^2\int_0^\infty
d{p}\;p^2\delta(E-p^2)\right]^{-1}\nonumber\\
&=& \left[\alpha+i2\pi^2 \sqrt{E}
\right]^{-1} \nonumber \\ &=& -(2\pi^2)^{-1}\left[-\frac{1}{
a^S}+\frac12r^S_0E-i \sqrt{E} \right]^{-1},
\end{eqnarray}
where $a^S$ and $r^S_0$ are the scattering length and effective
range, respectively, for the two-nucleon spin $S$. 

With the
ingredients above the kernel of the integrand becomes
\begin{eqnarray}
&&K^{(S)}({\bf p}^\prime_p,{\bf p}^\prime_n;{\bf
p}^{\prime\prime}_p,{\bf p}^{\prime\prime}_n)\nonumber\\
&=&\delta({\bf
p}^\prime_p-{\bf p}^{\prime\prime}_p)\delta({\bf p}^\prime_n-{\bf
p}^{\prime\prime}_n)  -(2\pi^2)^{-1}\delta\left({\bf
P}^{\prime\prime}_{pn,cm}-{\bf P}^\prime_{pn,cm}\right)
\nonumber
\\ &\times& \Bigg[ -\frac{1}{ a^S}+\frac12r^S_0\left(\Delta_n-\frac{{\hbar^2{\bf
P}^\prime}^2_{pn,cm}}{4m_N}\right)\nonumber\\
&-&i \sqrt{\Delta_n-\frac{{\hbar^2{\bf
P}^\prime}^2_{pn,cm}}{4m_N}}\Bigg]^{-1}
\nonumber \\
&\times&
\left[\Delta_n-\frac{({\hbar{\bf P}^{\prime\prime}_{pn,cm}})^2
}{4m_N}-\frac{({\hbar{\bf P}^{\prime\prime}_{pn,rel}})^2}{m_N}
+i\varepsilon\right]^{-1} .
\end{eqnarray}

In Ref. \cite{Bau09} the major findings were that the simple no-correlation, no-FSI calculation, does describe reasonably well
the qualitative aspects of the data, but fails when confronted with those of nn and pn final decay channels. In that reference one assumed the following: 1) plane waves for the outgoing protons and neutrons, and
2) no nn or pn interactions in the final channel, amounting to setting $\tau^{S}_{pn}(E_{pn}) = \tau^{S}_{nn}(E_{nn}) = 0$. Thus, one has,
\begin{equation}
\langle\chi^{(-)}({\bf p}_p)|{\bf
p}^\prime_p\rangle\langle\chi^{(-)}({\bf p}_n)|{\bf
p}^\prime_n\rangle= \delta({\bf p}_p - {\bf p}^\prime_p) \delta({\bf p}_n - {\bf p}^\prime_n), \label{EqApp1}
\end{equation}
\begin{equation}
K^{(S)}({\bf p}^\prime_p,{\bf p}^\prime_n;{\bf
p}^{\prime\prime}_p,{\bf p}^{\prime\prime}_n) = \delta({\bf p}^\prime_p - {\bf p}^{\prime\prime}_p) \delta({\bf p}^\prime_n - {\bf p}^{\prime\prime}_n), \label{EqApp2}
\end{equation}
and the formula for the decay width in Eq. (\ref{width}), reduces to 
\begin{eqnarray}
\Gamma&=& \frac{2\pi}{\hbar} \sum_{SM_SJ_FM_F} \int \left|\left<{\bf p}_n, {\bf p}_N, SM_SJ_FM_F |
V_{weak} |J_IM_I \right>\right|^2 \nonumber \\
&\times& \delta(E_{nN}+E_R-\Delta_N) \frac{d{\bf
p}_n}{(2\pi)^3}\frac{d{\bf p}_N}{(2\pi)^3}, \label{Old_Gamma}
\end{eqnarray}
which is the one also used in Ref. \cite{KGH10}.

\begin{figure}
\begin{center}
\resizebox{1.\columnwidth}{!}{
\includegraphics{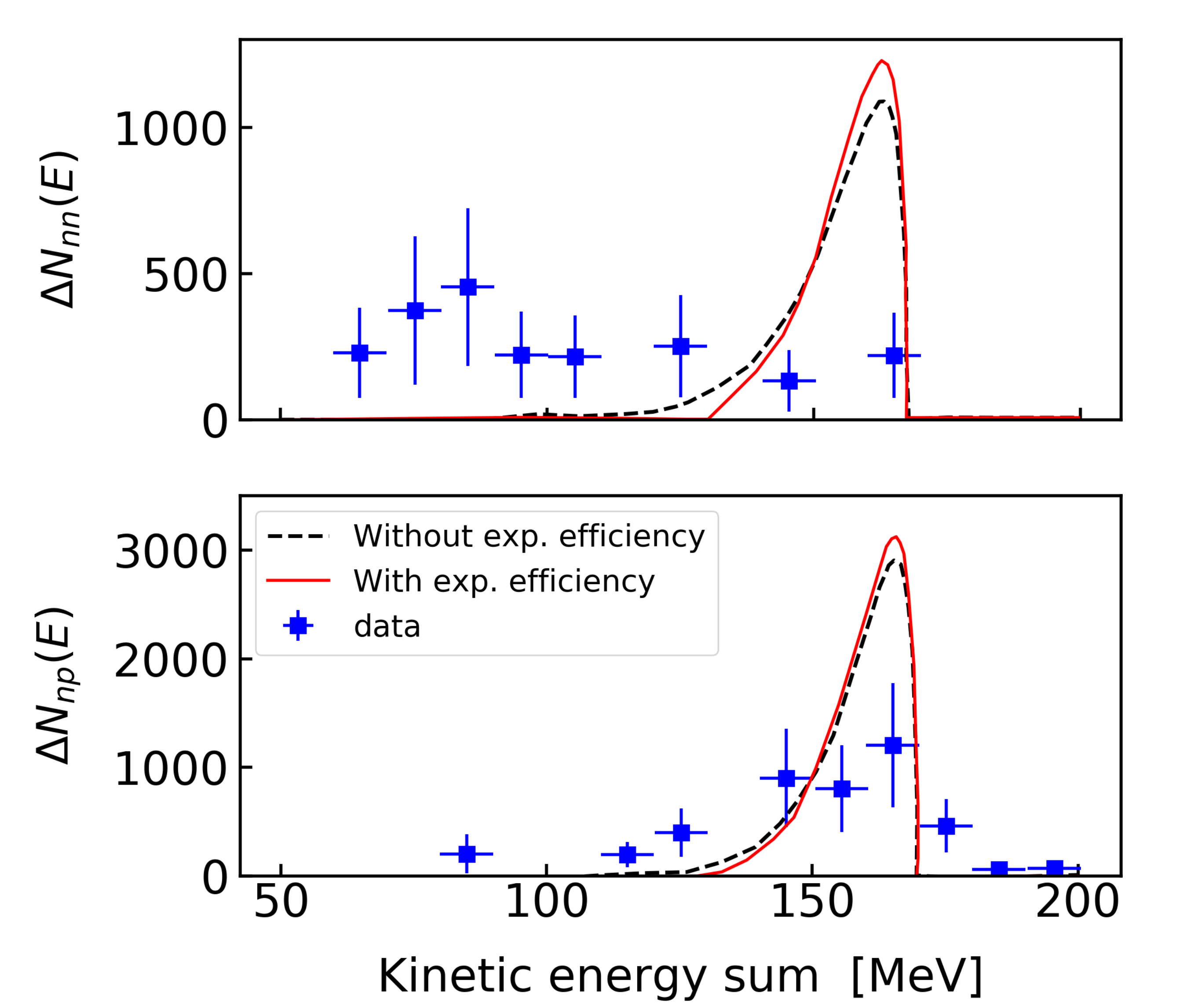}
}
\caption{The experimental \cite{Pa07} kinetic energy sums for proton-neutron  and for neutron-neutron pairs are compared to theoretical calculations with no FSI (see text) in the upper and in the lower panels, respectively. 
 The experimental data denoted by ${\Delta {\rm N}}_{nN}(E_i)$ are corrected for the detectors acceptance, neglecting events with  $E_N>25$ MeV or/and  scattering angles obeying $\cos\theta_{nN}<-0.5$.
The theoretical results using  Eq. (\ref{Old_Gamma}) with (solid curves) and without (dashed curves) the experimental cuts included in the calculations  \cite{Bau09}.}
\label{fig1} 
\end{center}
\end{figure}

The experimental kinetic energy sums of proton-neutron  and neutron-neutron pairs \cite{Pa07} are compared in Figure  \ref{fig1} (adapted from Ref. \cite{Bau09})  to theoretical calculations with no FSI in the upper and in the lower panels, respectively. 
 The experimental data denoted by ${\Delta {\rm N}}_{nN}(E_i)$ are corrected for the detectors acceptance, neglecting events with  $E_N>25$ MeV or/and  scattering angles obeying $\cos\theta_{nN}<-0.5$. The connection between the decay widths and the total kinetic energy of the emitted pairs, including details about the folding with detection efficiency is presented in Ref. \cite{KGH10}. 
 
 The theoretical results presented in Figure  \ref{fig1} use  Eq. (\ref{Old_Gamma}) with (solid curves) and without (dashed curves) the experimental cuts included in the calculations. The Figure  clearly shows that the low energy peaking in the spectrum calculated from $\Gamma$ is reasonable for the pn channel, while
it is quite bad for the nn channel. We attribute this to the approximations made above, Eq. (\ref{EqApp1}), and Eq. (\ref{EqApp2}). 

In the following, we present a new calculation of the spectra by including both the final nucleons'
distorted waves and the nucleon-nucleon interaction as explicit in Eq. (\ref{width}).\\

\section{Calculation details}

\subsection{An alternative formula for the width}
The weak decay process of $ A\to (A-2)+ N+N$ accounting for the
final state interaction can alternatively be calculated using the full Eq. (\ref{width})  with the account of FSI, where the nucleon optical
distortion by the remaining nucleus happens before the nucleon-nucleon final state interaction.  

We will present very general simplifying assumptions to obtain the distorted waves for each of the emitted nucleons which can be applied to the NMWD of any hypernucleus in practical calculations.
 
\subsection{FSI using eikonal waves}

The basic ingredients needed to calculate the width, Eq. (\ref{width}) are the distorted
waves of the outgoing nucleons and the two-nucleon t-matrix. For simplicity and insight
we use the eikonal approximation in the evaluation of the distorted waves. 

We take for the nucleon-nucleus optical potential the ``t$\rho$" approximation which relates the optical potential to phenomenological nucleon-nucleon scattering parameters
 \cite{HRB90}
\begin{equation}
V_{opt} (r, E) = t _{NN}(E) \rho_{A}(r),
\end{equation}
where $t_{NN}(E)$ is the average zero momentum transfer nucleon-nucleon t-matrix element,
and $\rho(r)$ is the matter density of the residual (assumed spherical) nucleus normalized as
\begin{equation}
4\pi \int_{0}^{\infty} r^2 dr \rho(r) = A,
\end{equation}
reproducing experimental quantities such as 
\begin{equation}
4\pi\int_{0}^{\infty}r^4 dr \rho(r) = A \left<r^2\right>,
\end{equation}
where $\left<r^2\right>$ is the squared rms radius of the residual nucleus. 

The distorted waves in the eikonal approximation
are given by
\begin{eqnarray}
\chi^{(-)}({\bf p}, {\bf r})^{\ast } &=& e^{-i {\bf k } \cdot {\bf r}} \nonumber \\
&\times& \exp{\left[-i\frac{k}{2E_p}\int_{z}^{\infty} dz' V_{opt}(\sqrt{z'^2 + b^2})\right]},\nonumber \\
\ 
\end{eqnarray}
which can be written as
\begin{equation}
\chi^{(-)}({\bf p}, {\bf r})^{\ast }= e^{-i {\bf k }\cdot {\bf r}} S(z, b),
\end{equation}
where the distortion factor (eikonal S-matrix) $S(z, b)$ has been introduced,
\begin{equation}
S(z, b) = \exp{\left[-i\frac{k}{2E}\int_{z}^{\infty} dz' V_{opt}(\sqrt{z'^2 + b^2})\right]}.
\end{equation}

In the following we assume for the matter density a Gaussian shape, $\rho(r) = \rho_{0}\exp{(-r^2/a^2)}$, which describes reasonable well the
light hypernuclei. Then, we have
\begin{equation} 
4\pi \int_{0}^{\infty} r^2 dr \rho(r) = A = \rho_{0} \pi^{3/2} a^3,
\end{equation} 
with
\begin{equation}
4\pi\int_{0}^{\infty}r^4 dr \rho(r) = A \left<r^2\right> = A R^2 = \rho_{0} \frac{3}{2}\pi^{3/2} a^5.
\end{equation}
Accordingly, we get the matter density at $r=0$ and $a$ in terms of the mass number and the radius of the residual nucleus,
\begin{equation}
\rho_{0} = \frac{A}{(\frac{2\pi}{3})^{3/2} R^3},
\end{equation}
and
\begin{equation}
a = \sqrt{\frac{2}{3}} R.
\end{equation}

Now we turn to the calculation of the distortion factor  $S(z, b)$, and use the Gaussian density in $V_{opt}$,
\begin{equation}
S(z, b) = \exp{\left[-i\frac{k V_{0}}{2E}\int_{z}^{\infty} dz' \exp{\left(-\frac{z'^2 + b^2}{a^2}\right)}\right]}.
\end{equation}
We write the above as,
\begin{equation}
S(z, b) = \exp{\left[-i\frac{k V_{0}}{2E}I(z, b)\right]}. \label{szb}
\end{equation}
The integral $$I(z, b) = \int_{z}^{\infty} dz' \exp{\left(-\frac{z'^2 + b^2}{a^2}\right)}$$ can be expressed in terms of the error function, $\Phi(z/a)$. That is,
\begin{equation}
I(z, b) = \exp{\left(-\frac{b^2}{a^2}\right)}\frac{\sqrt{\pi} a}{2}\left[1- \Phi\left(\frac{z}{a}\right)\right].
\end{equation}

The error function is evaluated from the incomplete Gaussian integral,
\begin{equation}
\Phi\left(\frac{z}{a}\right) = \frac{2}{\sqrt{\pi}}\int_{0}^{\frac{z}{a}}dx \ e^{-x^2}.
\end{equation}
The optical potential strength $V_{0} = \rho_{0} t_{NN}(E)$, is calculated from tabulated values of $t_{NN}(E)$ \cite{HRB90,AB20}. The imaginary
part of $t_{NN}(E)$ is directly related to the nucleon-nucleon total cross section as a consequence of the optical theorem.

The Fourier transform of $\chi^{(-)}({\bf p}, {\bf r})^{\ast }$ is needed in the calculation of the width. This is just
\begin{equation}
\chi^{(-)}({\bf p}, {\bf p'})^{\ast } = \frac{1}{(2\pi)^{2/3}}\int d{\bf r} e^{i{\bf k'}\cdot {\bf r}}e^{-i{\bf k} \cdot {\bf r}} S(z, b),
\end{equation}
where ${\bf p} = \hbar {\bf k}$. Using the cylindrical coordinates appropriate in eikonal-type calculation, we can reduce the above to
\begin{equation}
\chi^{(-)}({\bf p}, {\bf p'})^{\ast } = \frac{1}{(2\pi)^{3/2}} \int_{-\infty}^{\infty}dz e^{iq_{z}z}\int d{\bf b}e^{i {\bf q}_{b} \cdot {\bf b}} S(z, b),
\end{equation}
where we have introduced the momentum transfer ${\bf q} = {\bf k} - {\bf k'}$ and $d{\bf b} = bdbd\theta$. The product ${\bf q}_{b}\cdot{\bf b} = q_{b}b\cos\theta$. The integral over $\cos\theta$ results in a Bessel function, and 
the final expression for the distorted wave in momentum space is
\begin{equation}
\chi^{(-)}({\bf p}, {\bf p'})^{\ast } = \frac{1}{(2\pi)^{1/2}}\int_{-\infty}^{\infty}dze^{iq_{z}z}S(z, q_{b})      
\end{equation}
where $S(q_{b}, z)$ is given by
\begin{equation}
S(q_{b}, z) = \int_{0}^{\infty} bdbJ_{0}(q_{b}b)S(z,b).
\end{equation}

The perpendicular component of the momentum transfer, $q_{b}$, is much larger than the z-component, $q_z$. Accordingly, we evaluate the
b-integral above by setting $q_{b} = q$ and ignore the $e^{iq_{z}z}$ factor in the z-integral, resulting in
\begin{equation}
\chi^{(-)}({\bf p}, {\bf p'})^{\ast } = \frac{1}{(2\pi)^{1/2}}\int_{-\infty}^{\infty}dzS(z, q),  
\end{equation}
and
\begin{equation}
S(q, z) = \int_{0}^{\infty} bdbJ_{0}(qb)S(z,b).
\end{equation}
The above distorted wave in momentum space is to be compared to the momentum space representation of the plane wave which
is a 3-dimensional delta function $\delta({\bf q}) = \delta({\bf k}- {\bf k'})$.
The calculation of $S(q, z)$ requires $S(z, b)$ of Eq. (\ref{szb}).

The dependence of $S(q, z)$ on z is dictated by the integral 
$$\int_{0}^{\infty} bdbJ_0(qb)\exp{\left\{-i\alpha \exp[-b^2/a^2][1-\phi(z/a)]]\right\}},$$ 
with 
$\alpha = {k_p V_{0}}/{2E_p}$.
Since the Bessel function is a damped cosine function, one expects the function $S(q, z)$ to be a damped oscillatory 
function of $q$ as well.   For large value of $x$, $$J_{0}(x)= \sqrt{2/\pi x}\cos(x-\pi/4).$$ It is therefore convenient to write the Bessel function in terms of running waves of the form $e^{\pm iqb}$, and use the 
stationary point method to evaluate the resulting integrals. The stationary points, $b_{\pm}$, are obtained from
\begin{equation}
\pm q = -{d\over db}\left\{\alpha \exp\left(-b^2/a^2\right)[1-\phi(z/a)]\right\}_{\pm}.\label{stpha}
\end{equation}
The $b$-integrals can be evaluated by expanding the exponents $$\phi_{\pm} (q, b, z) = \pm qb - \alpha \exp\left(-b^2/a^2\right)[1-\phi(z/a)]$$ around the stationary points as
\begin{eqnarray}
\phi_{\pm}(q, b, z)&=&\pm qb_{\pm}-\alpha\exp\left(-b_{\pm}^2/a^2\right)[1-\phi(z/a)] \nonumber \\
&+& \frac{1}{2}\beta_{\pm} (b-b_{\pm})^2,
\end{eqnarray}
where $\beta$ is given by,
\begin{equation}
\beta_{\pm} = {d^2\over db^2}\left\{\alpha \exp\left(-b^2/a^2\right)[1-\phi(z/a)]\right\}_{\pm}. \label{betpm}
\end{equation}
The Gaussian $b$-integrals are readily performed using the result $$\int_{-\infty}^{\infty}b^{1/2}db e^{i\phi_{\pm}(q, b, z)} =\sqrt{i\pi b_{\pm}\beta_{\pm}/2}e^{i\phi(q, b_{\pm}, z)},$$ to give finally for the distorted wave factor $S(q, z)$,
\begin{eqnarray}
S(q, z)&=&\frac{1}{2q}\Big[\sqrt{b_{+}(z)\beta_{+}(z)}e^{i\phi(q, b_{+}, z)} \nonumber \\
&-&\sqrt{b_{-}(z)\beta_{-}(z)}e^{i\phi(q, b_{-}, z)}\Big],
\end{eqnarray}
where $b_{\pm}$ and $\beta_{\pm}$ are given by solving equations (\ref{stpha}) and (\ref{betpm}), respectively. When inserted in Eq. (45), the $z$-integral could be
evaluated through the use of the stationary point method too.

\subsection{FSI using the partial-waves method}
The outgoing proton and neutron distorted waves can also be calculated with the partial wave-method.
Usually this is more complicated than the eikonal waves, as one has to add many partial waves and numerically solve the Schr\"odinger equation. However, for the case of particle decay it is usually sufficient to  only care about s-waves for $\chi_{p,n}^{(-)}$, i.e.
\begin{equation}
\chi_{N}^{(-)}({\bf r})={1\over \sqrt{4\pi}}{u^{(-)}_N(r)\over r} ,
\end{equation}
where $N=p,n$ and we used $Y_{00}=1/\sqrt{4\pi}$ for the s-wave spherical harmonics.

The scattering waves $u^{(-)}$ are given by
\begin{equation}
u^{(-)}(r)=j_0(kr)+n_0(kr)\sin\delta_0,\label{s-wave}
\end{equation}
where $\delta_0$ is the s-wave phase-shift and $k=\sqrt{2E_N\mu_N}/\hbar$ is the relative momentum of the nucleon+residual nucleus, $\mu_N$ being their reduced mass.
Using the Born-approximation, the phase shift can be calculated from the N-nucleus potential with the formula
\begin{equation}
\delta_0=-{2\mu_N k\over \hbar^2}\int_0^\infty r^2 V_N(r) j_0^2(kr) dr.
\end{equation}

But notice that Eq. (\ref{s-wave}) is not very useful, as its Fourier transform diverges because of the second term. It is only valid in the asymptotic region. It would only work if the the central region of the wavefunction is suppressed. In fact, this occurs due to absorption. A simple way to care for absorption is to include an absorption radius, $R$, so that $u^{(-)}(r) =0$ for $r<R$ so that
\begin{equation}
\chi_{N}^{(-)}( p')=\int_R^\infty r^2 j_0(k'r)u^{(-)}(r) dr,
\end{equation}
where ${\bf p}'=\hbar {\bf k}'$. Another possibility is to use Eq. (\ref{szb}) in the form
\begin{equation}
\chi_{N}^{(-)}( p')=\int_0^\infty r^2 S(z=r,b=0) j_0(k'r)u^{(-)}(r) dr.
\end{equation}
In this case, the absorption is taken care by the eikonal S-matrix, $S$, and the integral can
start from the origin. The big advantage of this s-wave wavefunction is that its Fourier transform is angle independent.

\subsection{FSI using Migdal-Watson formula}
The  final state interaction (FSI) between neutrons has been worked out by Watson and Migdal \cite{Wat52,Mig54}. This idea has been used in later works to obtain the neutron-neutron scattering length \cite{Slo71} from, e.g., $^9$Be(n,nn)$^8$Be \cite{Bod90}, d + n $\rightarrow$ n  + n + p \cite{Breu74,Zei74,Gon99,Gon06}, $^2$H(n,np)n \cite{Huh00}, $K^-$ + d $\rightarrow$ $\bar K^0$ + 2n \cite{Coo64},   or $\gamma d \rightarrow \pi^+$nn \cite{Lens07}. These indirect methods are necessary, because direct nn scattering has not been carried out successfully yet. But the $^1S_0$ nn scattering length extracted from such indirect experiments range from approximately -16 fm to -19 fm.

In its standard form the Migdal-Watson model factorizes the reaction  production probability times a  FSI enhancement function, $F(E_{nN})$. As explained in Ref. \cite{Wat52} the matrix element describing a reaction with two-nucleons in the final channel can be factored into two parts at low energies, a part describing the production mechanism times the square of the relative wave function of the interacting pair, $\psi_{nN}({\bf k}, {\bf r}$),  averaged over the production region. If this wave function is normalized to unity for zero interaction, then it becomes an enhancement factor for the production process. 

Fermi \cite{Fer51} suggested a suitable
normalized enhancement factor to be the square of the ratio of the wave function $\psi_{nN}({\bf k}, {\bf r}$)   to the wave function corresponding to zero phase shift, the ratio being evaluated at a radius corresponding to the range of the interaction, $r_{nN}$. If one uses the asymptotic wave function for s-wave continuum states and the effective range expansion, the enhancement factor $F(E_{nN})$ for an nN final-state interaction is \cite{Boy69}
\begin{eqnarray} 
F(E_{nN}) &=& \left| {\psi(k_{nN},r_{nN}) \over \psi^{(0)}(k_{nN},r_{nN})}\right|^2 \nonumber \\
&=&{(1/r_{nN}-1/a_{nN} +k_{nN}^2r_{nN}/2)^2 \over (-1/a_{nN} + k_{nN}^2r_{nN}/2)^2 +k_{nN}^2},
\label{Gnn}
\end{eqnarray}
where $E_{nN}$ and $\hbar k_{nN}$ denote the relative energy and the relative momentum of the nN pair, respectively.

Using this method, the plane wave prescription  as described by Eq. (\ref{Old_Gamma}) is modified by including the FSI in its integrand. Therefore, using the Migdal-Watson (MW) model, the decay width becomes equal to
\begin{eqnarray}
\Gamma_N^{MW}&=& \frac{2\pi}{\hbar} \sum_{M_SJ_FM_F} \int \left|\left<{\bf p}_n, {\bf p}_N, SM_SJ_FM_F |
V_{weak} |J_IM_I \right>\right|^2 \nonumber \\
&\times& F(E_{nN})\delta(E_{nN}+E_R-\Delta_N) \frac{d{\bf
p}_n}{(2\pi)^3}\frac{d{\bf p}_N}{(2\pi)^3}. \label{New_Gamma}
\end{eqnarray}

The singlet $^1S_0$ scattering length and the effective-range for neutron-neutron scattering has been determined indirectly and is assumed to be $a_{nn}^{s} = -18.9 \pm 0.4$ fm and $r_{nn}^{s}=2.819$ fm \cite{MS01,Gard09}, although much lower  values for the scattering length (as much as 17\%) have been reported in the literature \cite{Huhn00}. 

The neutron-neutron zero values triplet scattering parameters are unknown. In principle, they could be determined by a crossed beam experiment of polarized neutrons, but the presently available intensities  of neutron sources are too weak to allow for such experiments. 

For a neutron-proton pair both the triplet and singlet zero value scattering parameters are known, e.g., $a^t_{np} = 5.4114 (27)$ fm, $r_{np}^t = 1.7606  (35)$ fm, $a^s_{np} =-23.7154(80)$ fm and $r^s_{np} = 2.706 (67)$ fm \cite{BP10}. 

\begin{figure}[t]
\begin{center}
\resizebox{1.\columnwidth}{!}{
\includegraphics{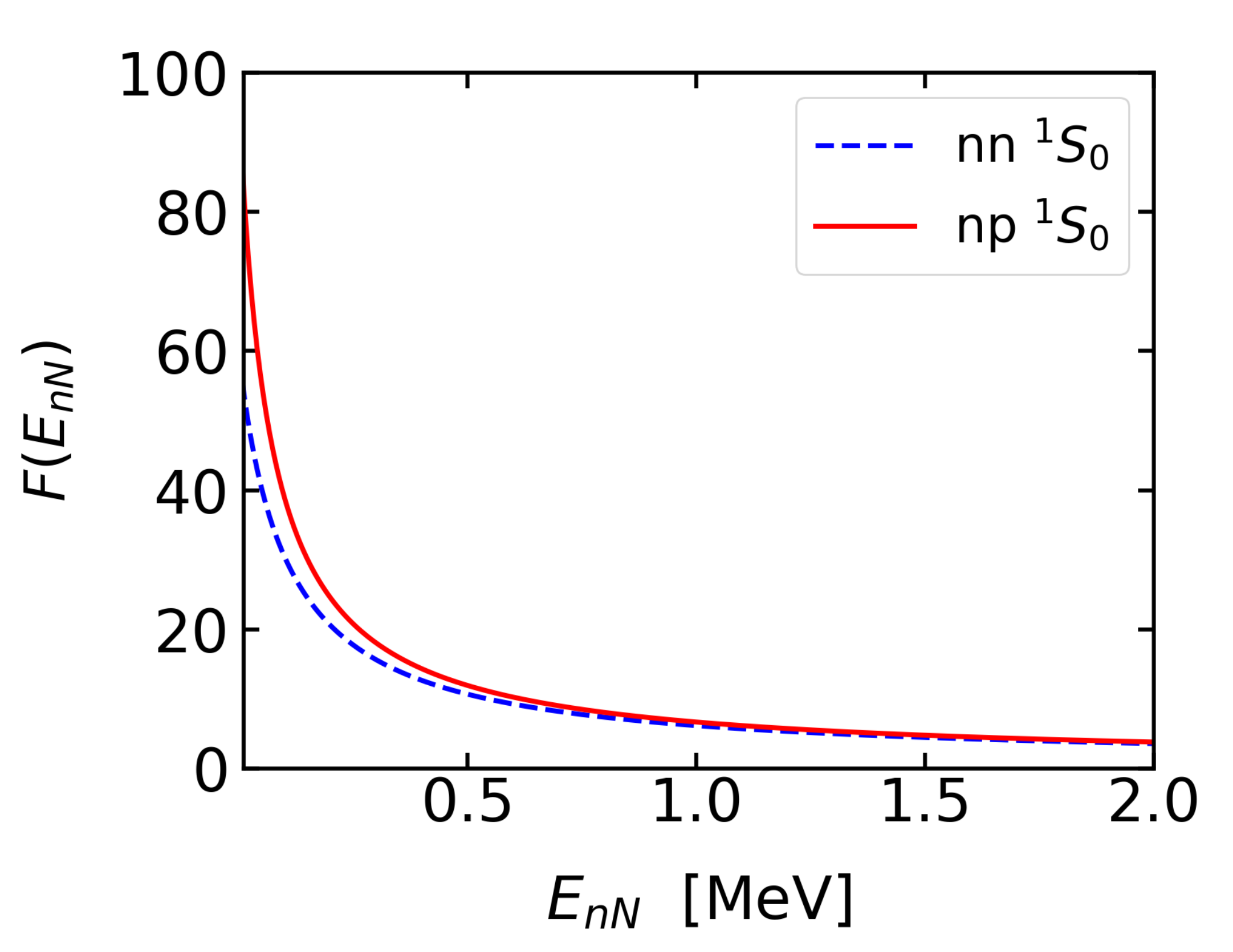}
}
\caption{The enhancement function of Eq. (\ref{Gnn}) for the $^1S_0$ singlet neutron-neutron (nn, dashed line) and neutron-proton (np, solid line) systems.}
\label{FE} 
\end{center}
\end{figure}

In Figure  \ref{FE} we show the enhancement function based on Eq. (\ref{Gnn}) for the $^1S_0$ singlet neutron-neutron (nn) and neutron-proton (np) systems. It is evident that the main influence of the FSI due to the nucleon-nucleon interaction occurs at very low relative energy of the pair. Therefore, it is clear that this kind of FSI will have little effect on the observed kinetic energy sums for proton-neutron and for neutron-neutron pair shown in Figure  \ref{fig1}. However, it might be useful to determine the low relative energy spectrum of the pair. With the increased efficiency of neutron detectors, experiments aimed at studying the nn scattering length by detection of high energy neutrons are a real possibility to improve the accuracy of the nn scattering lentgth \cite{Aum20}.

\subsection{FSI with spin correlations}
The effects of spin correlations also play an important role in final state interactions for the emission of nucleon pairs from a nucleus. The effect of spin correlations on the spectra of two-nucleon decays have been studied extensively. Here we will borrow some fo the findings presented in Ref.  \cite{BHV08} where the interesting aspect of  nuclear entanglement was explored. For the case of NMWD spin correlations affect the two-neutron decay mode. 

Being identical particles, the detection of the two neutrons  requires the consideration of their admixture of singlet and triplet states.  When a  single neutron is described  initially by a localized wave-packet $\psi_0({\bf r}_1)$,  the probability amplitude to detect it
with momentum ${\bf p}_1$ is given by 
\begin{equation}
{\cal A}({\bf p}_1, {\bf r}_1) =\int d^3 r \chi^{(+)}({\bf p}_1, {\bf r})K({\bf r}, {\bf r}_1)\psi_0({\bf r}_1),
\end{equation}
where  $\chi^{(+)}({\bf p}_1, {\bf r})$ is an asymptotic outgoing wave with energy  $E_1=p_1^2/2M$ and $K({\bf r}, {\bf r}_1)\psi_0({\bf r}_1)$ is the propagator accounting for the wavefunction evolution from the source to the asymptotic region.

Without consideration of spin correlations, when the neutron 1 is detected with momentum ${\bf p}_1$ and neutron 2 with momentum ${\bf p}_2$, the probability amplitude for this event is the product ${\cal A}({\bf p}_1, {\bf r}_1){\cal A}({\bf p}_2, {\bf r}_2)$. Since the particles are indistinguishable, for a singlet (triplet) state $S=0$ ($S=1$) this probability amplitude must be symmetric (antisymmetric) and the normalized probability amplitude reads
\begin{eqnarray}
{\Lambda}^{(\pm)}({\bf p}_1, {\bf p}_2, {\bf r}_1, {\bf r}_2) &=&{1\over \sqrt{2}}\Big[
{\cal A}({\bf p}_1, {\bf r}_1){\cal A}({\bf p}_2, {\bf r}_2) \nonumber \\
&\pm& {\cal A}({\bf p}_2, {\bf r}_1){\cal A}({\bf p}_1, {\bf r}_2 \Big],
\end{eqnarray}
with the plus (minus) sign for the singlet (triplet) spin-state. 

In Ref. \cite{BHV08} the two-particle momentum probability to measure a neutron with momentum ${\bf p}_1$ in coincidence with another with momentum ${\bf p}_2$, was defined as
\begin{eqnarray}
P({\bf p}_1, {\bf p}_2) &=&\int d^3r_1d^3r_2\Bigg| \Big|
{\Lambda}^{(+)}({\bf p}_1, {\bf p}_2, {\bf r}_1, {\bf r}_2)\Big|^2\nonumber \\
&\pm& {\cal M} {\Lambda}^{(-)}({\bf p}_1, {\bf p}_2, {\bf r}_1, {\bf r}_2)\Big|^2 \Bigg|^2 \Big], \label{pp1p2}
\end{eqnarray}
where $\cal M$ is a mixing parameter which parametrizes the triplet state relative contribution. 

A correlation function
\begin{eqnarray}
C({\bf p}_1, {\bf p}_2) = {P({\bf p}_1, {\bf p}_2) \over P({\bf p}_1) P({\bf p}_2)}
\end{eqnarray}
can be easily built from Eq. (\ref{pp1p2}) to determine the relative contribution of the singlet and the triplet states in the initial configuration of the neutron pair. In this equation, $P({\bf p}_1)$ and $P({\bf p}_2)$ are the probabilities of observing ${\bf p}_1$ and ${\bf p}_2$ independently, with no coincidence measurements. As shown in Ref. \cite{BHV08}, the impact of spin admixtures in the correlation function are  larger for relative momenta $|{\bf p}_1 - {\bf p}_2|/2 \sim 20$ MeV/c, based on the typical localization ranges  $r\lesssim 4$ fm of the pair within the nucleus. As expected, the correlation  is sensitive to the  admixture of triplet and singlet states..

\begin{figure}[t]
\begin{center}
\resizebox{1.\columnwidth}{!}{
\includegraphics{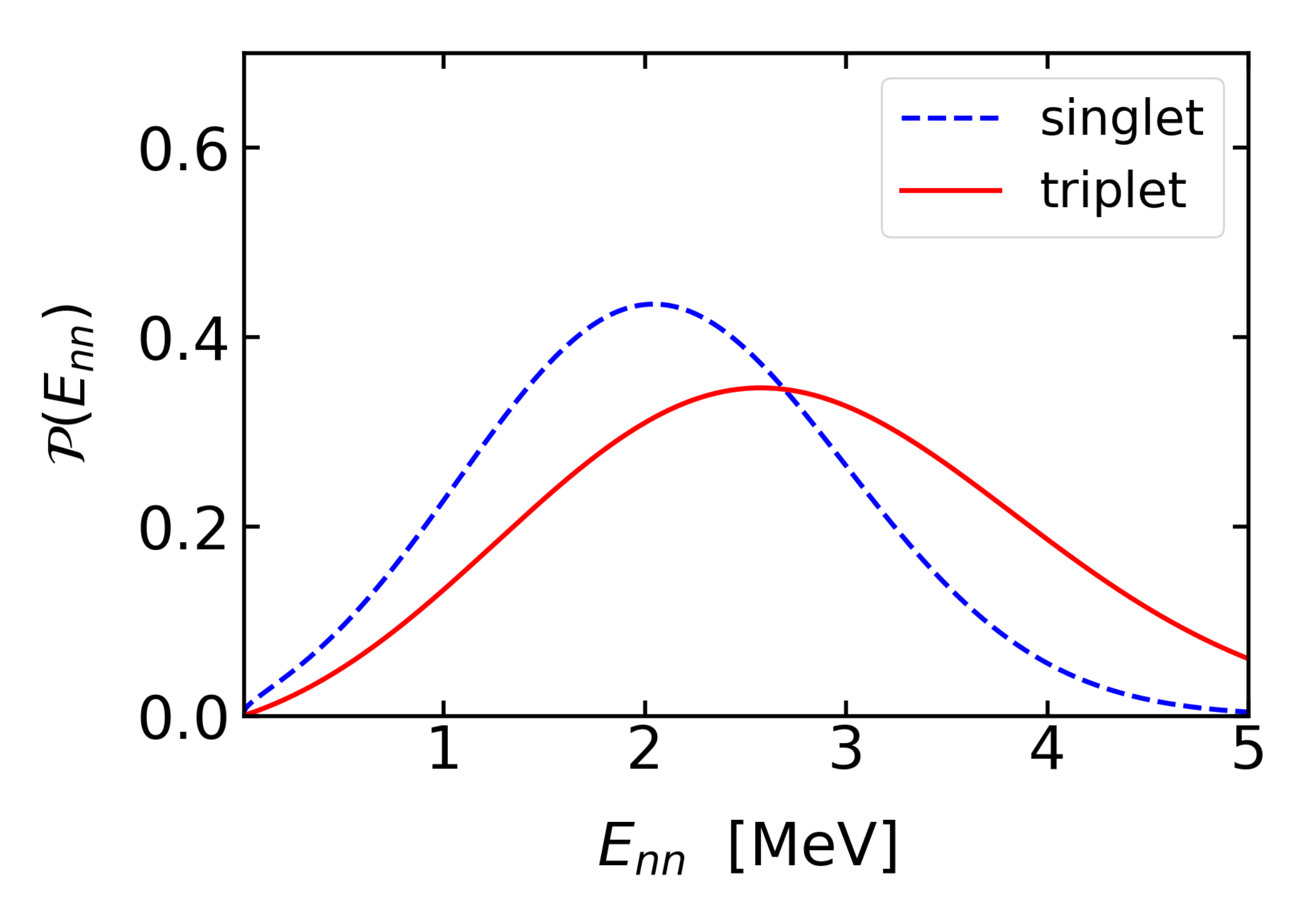}
}
\caption{Momentum density as a function of the relative energy of the nn pair. The dashed (solid) curve shows results for singlet (triplet) states.}
\label{pnp} 
\end{center}
\end{figure}

Figure  \ref{pnp} shows the results for the two-body probability density ${\cal P}(E_{nn})$, obtained from Eq. (\ref{pp1p2}) so that $P(E_{nn})={\cal P}(E_{nn})dE_{nn}$, as a function of the relative energy of the nn pair. The dashed (solid) curve shows results for singlet (triplet) states. The function $P({\bf p}_n, {\bf p}_{n'})$ was calculated using the asymptotic expansion of the neutron wavefunction depending on the the zero value scattering parameters.  We arbitrarily assumed $r_0 = 4$ fm for the average initial distance of the nn pair right after the NMWD. 

It is clear from Figure  \ref{pnp}  that spin correlations depend on the size of the source function and  the they spread over a larger relative energy of the pair than the final state interaction described in the previous section. Thus the initial configuration of the emitted pair right after the NMWD and the symmetry of the nn wavefucniton plays a larger role in the kinetic energy sum of neutron-neutron pair than the final state interaction discussed in the context of Figure  \ref{FE}.

Within the same premises as those used in the derivation of Eq. (\ref{New_Gamma}), the effect of spin correlation in the decay width can be obtained by folding the correlation probability amplitude of Eq. (\ref{pp1p2}) with the decay formula (\ref{Old_Gamma}). That is, inclusion of spin correlations (sc) modify the decay width formula to 
\begin{eqnarray}
\Gamma_N^{sc}&=& \frac{2\pi}{\hbar} \sum_{M_SJ_FM_F} \int \left|\left<{\bf p}_n, {\bf p}_{n'}, SM_SJ_FM_F |
V_{weak} |J_IM_I \right>\right|^2 \nonumber \\
&\times& P({\bf p}_n, {\bf p}_{n'}) \delta(E_{nn'}+E_R-\Delta_{n'}) \frac{d{\bf
p}_n}{(2\pi)^3}\frac{d{\bf p}_{n'}}{(2\pi)^3}. \label{New_Gamma2}
\end{eqnarray}

\section{Contribution of FSI}

\begin{figure}[t]
\begin{center}
\resizebox{1.\columnwidth}{!}{
\includegraphics{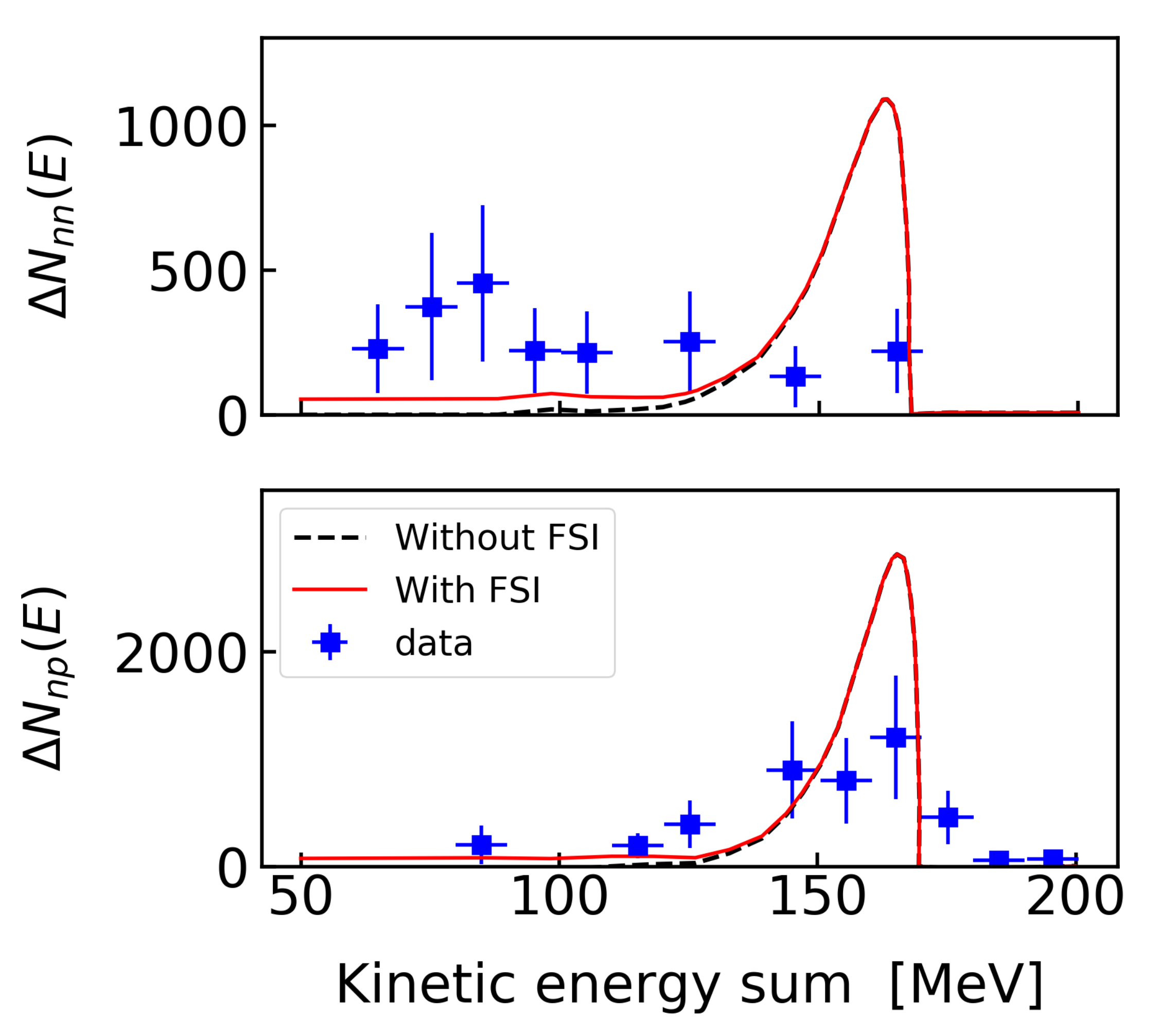}
}
\caption{The enhancement function of Eq. (\ref{Gnn}) for the $^1S_0$ singlet neutron-neutron (nn) and neutron-proton (np) systems.}
\label{FSI} 
\end{center}
\end{figure}

We now consider the contributions of all FSI following the calculation details discussed in the previous section. This is shown in Figure  \ref{FSI} where the dashed curves are corrected for the detectors acceptance, as described previously. The solid curve in Figure  \ref{FSI} contains the changes due to FSI. 

One sees from Figure  \ref{FSI} that the FSI additional contributions to the calculitions obtained with plane waves are not of major relevance. The effects of FSI in the calculations are only visible at low total kinetic energy of the pair. The largest FSI contribution arises from considering distorted waves for the emitted nucleons, as described in Section 4. The effects of final state interaction and the spin-correlations (assuming an arbitrary  share of singlet and triplet states) are not enough to modify the spectrum appreciably.

\section{Conclusions}

In conclusion, we derived a useful formalism to describe final state interactions in the NMWD of a hypernucleus. As an application and comparison with experimental data, we have supplied an estimate of the corrections due to final state interactions in the non-mesonic decay of $_\Lambda^4$He. Evidently, our approach can be used for the NMWD of any other nucleus. 

Our calculations are compared to experimental data on the sum of the kinetic energies of neutron-neutron and neutron-proton pairs. We considered the effects of (a) distortions caused by the FSI of the outgoing nucleus with the residual nucleus, (b) the FSI due to the relative interaction between the outgoing pair, and (c) the effect of quantum symmetries on spin correlations of the emitted pair in their source function just after the day. It was concluded that the later two contributions are small while the distortion caused by interaction with the residual nucleus causes an enhancement in the spectrum at lower kinetic energies, although not enough to display an agreement with the measurements for the NMWD of $^4_\Lambda$He. 

It is necessary to highlight some limitations of our study: (a) In our three-body model, we include final state interactions between the nucleons and the residual nuclei, as well as between the two nucleons. However, we do not include ``hard'' binary collisions between the emitted nucleons and  the  nucleons of the residual nucleus. This could be the reason why we see a small effect of the final state interaction, although we expect the effect to be small for $^4$He, with only two other nucleons to collide with. (2) We also make it clear that $^4_{\Lambda}$He decays mostly (80\% branching ratio) via mesonic modes, and only one-fifth of the time via pn and nn emission, and a very tiny number of events occur through the decay of three nucleons. Additionally, non-mesonic two-body (80\%) and three-body (20\%) modes dominate in heavier nuclei like $^{12}_{\Lambda}$C, not considered here. An extension of our studies to NMWD including binary nucleon-nucleon collisions and the complications  emerging for heavier nuclei is warranted. Work in this direction is in progress \cite{BL21}.

\section*{Acknowledgements} 
We have benefited from useful discussions with M.S. Hussein,  T. Frederico and F. Krmpotic, C. De Conti and T. Aumann. This work was partially supported by the U.S. Department of Energy under contract No. DE-FG02-08ER41533 and funding contributed through the LANL Collaborative Research Program by the Texas A\&M System National Laboratory Office and Los Alamos National Laboratory.
%\end{acknowledgement}


\begin{thebibliography}{99}
\section*{REFERENCES}
\bibitem{DP53} M. Danysz, J. Pniewski,  Bull. Acad. Pol. Sci. 1, 42 (1953); M. Danysz, J. Pniewski, Phil. Mag. 44, 348 (1953).
\bibitem{Ams08} C. Amsler et al., Phys. Lett. B 667, 1 (2008). 
\bibitem{Oku82} L.B. Okun, Leptons and Quarks (North-Holland, Amsterdam, 1982).
\bibitem{Isg90} N. Isgur, K. Maltman, J. Weinstein, and T. Barnes, Phys. Rev. Lett. 64, 161 (1990).
\bibitem{GHM16} A. Gal, E.V. Hungerford, and D.J. Millener, Rev. Mod. Phys. 88, 035004 (2016).
\bibitem{Bar08} C. Barbero, A.P. Galeao, M.S. Hussein, F. Krmpotic, Phys. Rev. C 78, 044312 (2008).
\bibitem{Bau09} E. Bauer, A. P. Galeao, M. S. Hussein, F. Krmpotic, J. D. Parker, Phys. Lett. B 674, 103 (2009).
\bibitem{Bau09b} E. Bauer, A. P. Galeao, M. S. Hussein, F. Krmpotic, J. Parker, Phys. Lett.. B 674,  103 (2009).
\bibitem{KGH10} F. Krmpotic, A. P. Galeao, and M. S. Hussein, AIP Conference Proceedings 1245, 51 (2010).
\bibitem{Baue10} E. Bauer, A. P. Galeao, M.S. Hussein, F. Krmpotic,   Nucl. Phys. A  834, 599c  (2010).
\bibitem{BHV08} C. A. Bertulani, M. S. Hussein and G. Verde, Phys. Lett. B 666, 86 (2008).
\bibitem{Pa07} J.D. Parker, et al., Phys. Rev.  C 76, 035501 (2007).
\bibitem{HRB90} M. S. Hussein, R. A. Rego, and C. A. Bertulani, Phys. Reports 201, 279 (1991). 
\bibitem{AB20} T. Aumann and C. A. Bertulani, Prog. Part. Nucl. Phys. 112, 103753 (2020).
\bibitem{Parr02} A. Parr\~{n}o and A. Ramos,  Phys. Rev. C 65, 015204 (2002).
\bibitem{Ito08} K. Itonaga, T. Motoba, T. Ueda, and Th. A. Rijken, Phys. Rev. C 77, 044605 (2008).
\bibitem{Wat52} K. M. Watson,  Phys. Rev. 88, 1163 (1952). 
\bibitem{Mig54} A. B. Migdal,  JETP (Sov. Phys.) 1, 2 (1954).
\bibitem{Slo71} R. J. Slobodrian, Rep. Prog. Phys.  34, 175 (1971).
\bibitem{Bod90} K. Bodek et al., Few-Body Systems 8, 23 (1990).
\bibitem{Breu74}W. H. Breunlich, S. Tagesen, W. Bertl and A. Chalupka, Nucl. Phys. A 221, 269 (1974).
\bibitem{Zei74} B. Zeitnitz, et al., Nucl. Phys. A 231, 13 (1974).
\bibitem{Gon99} D.E. Gonzalez Trotter et al., Phys. Rev. Lett. 83, 3788 (1999).
\bibitem{Gon06} D.E. Gonzalez Trotter et al., Phys. Rev. C 73, 034001 (2006).
\bibitem{Huh00} V. Huhn et al., Phys. Rev. C 63, 014003 (2000).
\bibitem{Coo64} P. A. Cook, Lett. Nuovo Cimento 1, 1364 (1964).
\bibitem{Lens07} V. Lensky, et al., Eur. Phys. J. A 33, 339 (2007).
\bibitem{Fer51} E. Fermi, ``Elementary Particles" (Yale University Press, New Haven, Conn., 1951), p. 58-64.
\bibitem{Boy69} R. J. Slobodrian, Rep. Prog. Phys. 34, 175 (1971).
\bibitem{MS01} R. Machleidt and I. Slaus, J. Phys. G: Nucl. Part. Phys. 27, R69 (2001).
\bibitem{Gard09} Anders Gardestig, J. Phys. G: Nucl. Part. Phys. 36,  053001 (2009).
\bibitem{Huhn00} V. Huhn et al., Phys. Rev. Lett. 85, 1190 (2000); Phys. Rev. C63, 014003 (2000).
\bibitem{BP10} V.A. Babenko and N.M. Petrov, Phys. Atom. Nucl. 73,  1499 (2010).
\bibitem{Aum20} T. Aumann, private communication.
\bibitem{BL21} C. A. Bertulani and R. Lobato, in preparation.
\end{thebibliography}
\end{document}